\documentclass[11pt]{article}

\usepackage[top=1in,bottom=1in,left=1in,right=1in]{geometry}
\usepackage[english]{babel}
\usepackage[utf8]{inputenc}
\usepackage[T1]{fontenc}
\usepackage{lmodern}
\usepackage{graphicx}
\usepackage{tikz}
\usetikzlibrary{calc}
\usepackage{pgfplots}
\pgfplotsset{compat=newest}
\usepgfplotslibrary{groupplots}
\usepackage{xcolor}
\usepackage{amsmath}
\usepackage{amsfonts}
\usepackage{mathtools}
\usepackage{amssymb}
\usepackage{accents}
\usepackage{algorithm}
\usepackage{algpseudocode}
\usepackage{natbib}
\usepackage{hyperref}
\usepackage{float}
\usepackage{caption}
\usepackage{subcaption}
\usepackage[font=small, labelfont=bf, margin=0.25in]{caption}

\addto\extrasenglish{%

}

\begin{document}

\title{Cheap Talking Algorithms\thanks{We thank Fran\c{c}oise Forges, Alkis Georgiadis-Harris, Balazs Szentes, ChatGPT4, and seminar participants at the University of Warwick and at the 2023 Bergamo SkIO conference for their valuable comments. We are also grateful to Sophia Skenderis for her research assistance and to several anonymous reviewers. All computations have been performed using the Julia programming language. The code to replicate the results is available at \url{https://github.com/massimilianofurlan/rl_cheap_talk}.}}
\date{\today}
\author{Daniele Condorelli\thanks{Department of Economics, University of Warwick, UK. Email: d.condorelli@warwick.ac.uk.} \hspace{3cm} Massimiliano Furlan\thanks{Department of Economics, University of Warwick, UK. Email: massimiliano.furlan@warwick.ac.uk.}
}
\maketitle

\begin{abstract}
We simulate behaviour of two independent reinforcement learning algorithms playing the \citet{CrawfordSobel82} game of strategic information transmission. We adopt memoryless algorithms to capture learning in a static game where a large population interacts anonymously. We show that sender and receiver converge to Nash equilibrium play. The level of informativeness of the sender's cheap talk decreases as the bias increases and, at intermediate level of the bias, it matches the level predicted by the Pareto optimal equilibrium or by the second best one. Conclusions are robust to alternative specifications of the learning hyperparameters and of the game.
\end{abstract}

	
\section{Introduction}
Consider the classic signalling game: a sender is informed about a payoff-relevant parameter drawn from a known distribution and takes one of several possible actions; an uninformed receiver observes the sender's action but not the parameter, and makes a decision. In a landmark article, \citet{CrawfordSobel82}  (henceforth CS) showed that, even if the payoffs of both agents are independent of the sender's action, there are equilibria where the action transmits information about the parameter, as long as the conflict of interest about the ideal receiver's decision is not too large. By interpreting the payoff-irrelevant actions of the sender as ``cheap talk'', CS delivers a powerful formal theory of communication. Non-committal and purely symbolic behaviour can convey information and help coordinate subsequent interactions even if rational agents do not share identical goals.

In this paper, we compute stationary points of memoryless independent reinforcement learning algorithms playing the CS's game of information transmission.\footnote{Computational techniques are necessary because finding limit points of independent learning algorithms interacting with each other is, to date, an intractable mathematical problem. Methods relying on approximation via systems of differential equations are not directly applicable to our case (see \citet{BorgersSarin97} and \citet{BanchioMantegazza22}).} These algorithms work roughly as follows. For each of a finite set of types, the sender keeps track of a vector, which stores its current estimates of the value of taking each action given the type. The receiver, instead, holds a vector for each of the signals the sender may send. Any such vector contains the receiver's estimate of the value of each action following a given signal. In each period, the algorithms select actions following a softmax policy. Most likely, they take the highest-reward action according to their estimates, but with some probability they experiment with other actions. Such probability decays over time, depending on a hyperparameter (i.e., the temperature-decay factor). After both agents have moved, the relevant estimates are updated to account for the payoffs received. Another hyperparameter (i.e., the learning rate) establishes how much the current experience is weighted vis-a-vis the past.\footnote{The reinforcement learning literature has proposed numerous learning algorithms. We chose one of the simplest forms of reinforcement learning available for our environment \citep{SuttonBarto18}. } 

We find that a sender and a receiver playing together in repeated instances of the CS's game converge to Nash equilibrium behaviour with substantial information transmission, which grows monotonically as the bias decreases.  Except at levels of the bias that make perfect information transmission an equilibrium, where behaviour is more nuanced, the mutual information between the distribution of the type and that of the message chosen by the sender (i.e., the informativeness of the sender's cheap talk) matches that of the maximally informative, Pareto optimal, equilibrium or of the second most informative one. As the bias grows, the sender's strict preference for the Pareto optimal equilibrium attenuates and, when near indifference is reached, information transmission drops to the second-best equilibrium, which itself becomes the optimal one once the bias increases~further.

Our algorithms play a very large number of identical one-shot games together. However, at convergence, play is independent of past choices. By design, algorithms are learning to play a static game, not a dynamic one. Therefore, the reader may wonder why not endowing them with the ability to respond to past history, as this allows for a richer set of outcomes and, quite possibly, more communication in the long run. While this would be a valid modelling choice were we interested in the emergence of bilateral communication, it is not the right approach to evaluate the emergence of language in a large population, which is the focus of our analysis and, arguably, the appropriate interpretation of equilibrium in CS. In fact, building on a standard, if not entirely uncontroversial argument in the theory of learning in games, our results should be interpreted as arising from learning within a large population of possibly more sophisticated reinforcers interacting anonymously (see \citet{FudenbergKreps93} and \citet{FudenbergLevine98}).

Our main contribution is to the theory of strategic information transmission. Following pioneering work in psychology (e.g., \citet{BushMosteller55}) and in game theory (e.g., \citet{ErevRoth98}), reinforcement learning offers a model of human behaviour alternative to the traditional game-theoretic one.\footnote{\citet{ErevRoth98} wrote: ``well-developed, cognitively informed adaptive game theory will complement conventional game theory, both as a theoretical tool and as a tool of applied economics.'' However, in contrast to what was hoped for by \citet{ErevRoth98}, simple reinforcement learning algorithms do not fit the experimental data well in this case. In fact, over-communication relative to the most informative equilibrium is a common feature of experimental implementations of the CS game involving human subjects. (see \citet{DickhautEtal95} and, especially, \citet{CaiWang06}).} In this light, our results complement the existing knowledge in two ways. First, we show that information transmission in cheap talk games with conflict of interest is a robust feature of play, emerging from a different modelling approach to strategic interaction. To our knowledge, theoretical results on the convergence of reinforcement learning algorithms to informative equilibria are only available in the case where agents have aligned interests. Both \citet{YileiEtAl2011} and \citet{Wrneryd1993} demonstrate convergence of forms of reinforcement learning algorithms to the most informative equilibrium. Second, we contribute to the large game-theoretic literature on equilibrium selection in games with information transmission. As it is well known, such games possess many qualitatively different equilibria. Our main result delivers a cautionary tale regarding the consensus reached in the uniform-quadratic environment around the selection of the most informative and Pareto optimal equilibrium. In particular, in addition to convergence to the first best, we also show convergence to the second most informative equilibrium at certain levels of the bias, where such an equilibrium is also close to optimal for the sender.\footnote{For an important recent contribution to the equilibrium selection literature see \citet{ChenEtal08}. Most closely related to our work is perhaps the evolutionary and learning approach to selection, which shows that the most informative equilibrium is the evolutionarily stable outcome of the CS game when it exists (see \citet{BlumeEtal93}) and the limit point of the best-response dynamics (see \citet{GordonEtal22, SemiratForges24}).} 

While experimental evidence shows that communication in cheap talk games with partial conflict of interests is achieved by humans (e.g., see \citet{BlumeEtal20} for a survey), to our knowledge an analogous conclusion has not yet been robustly established for artificially intelligent agents (AI agents). Most of the machine learning literature has focused on games with common interest, observing that AI agents learn to communicate successfully (e.g., see \citet{LazaridouEtal16}, \citet{HavrylovTitov17}, \citet{FoersterEtal16}). Instead, mostly negative results have been obtained in games where agents have conflicting interests (e.g., see \citet{CatEtal18}). An important exception is  \citet{NoukhovitchEtal21}. They consider a CS game played on a circle, for which equilibrium characterisation is not available. Employing AI agents controlled by neural networks they show that communication is achieved even when the bias of the sender is non-zero. We depart from \citet{NoukhovitchEtal21} by employing simple reinforcement learners and by looking at the original (discretised) CS game. Doing this allows us to compare simulation outcomes to the theoretical benchmark and establish that communication often takes place at the highest level predicted by theory even when a very simple model of learning is adopted.

Finally, we hope that the observation that private information can be successfully communicated between AI agents will open up new questions within a growing literature in economics which, motivated by policy concerns, looks at AI agents playing various market games. Contributions to this recent literature include \citet{CalvanoEtal20}, \citet{BanchioSkrzypacz22}, \citet{AskerEtal22}, \citet{Johnson2023} and \citet{DeCarolisEtal23}.\footnote{The literature on market games played by AI agents was initiated by computer scientists, with early contributions including \citet{WaltmanKaymak08} and \citet{TesauroKephart02} among others.} A central theme of this research agenda is showing that AI agents learn to play strategies that deliver supra-equilibrium profits, which would be deemed implicitly collusive if played by humans. 
Since communication expands the equilibrium set in a game-theoretic sense (e.g., see \hbox{\citet{AumannHart03}}), the possibility of communication raises the question of what should be expected in market games played by algorithms if collusion can be explicit. This is not a moot concern, even when a direct communication channel is not part of market design. In fact, as auction practice has shown, bidders learn to exchange information in very imaginative ways, for instance by using the last digits of their submitted bids.\footnote{\citet{BajariYeo09} suggest that in some FCC spectrum auctions bidders used such form of code-bidding to communicate their intentions and avoid competing on the same portions of the spectrum for sale.} Since we expect sophisticated AI agents to exploit all communication opportunities, our results suggest that explicit collusion between algorithms with a sufficiently large state space and a long history of interactions may be as worrisome as the implicit one uncovered by the existing literature.

In the next section, we present the discretised uniform-quadratic specification of the CS game and our simulation design. \autoref{sec:results} presents the main results. In \autoref{sec:robustness}  we illustrate the robustness of our findings in terms of the parameters of the game. \autoref{sec:conclusion} concludes with some avenues for future work.

\section{RL agents playing the cheap talk game}\label{sec:rl_cheap_talk}

There are two agents, a sender ($S$) and a receiver ($R$). At the outset, a type  $\theta$ is drawn from a known uniform distribution with PDF $p$ and support over a finite set $\Theta$, which is composed by $n$ uniformly spaced points in the interval $[0,1]$. The sender privately observes the realised type $\theta$ and sends a message $m \in M$ to the receiver, with $|M| = | \Theta |$. Then, the receiver observes message $m$ and takes an action $a \in A$, with $A$ formed by $2n - 1$ uniformly spaced points in $[0,1]$. The receiver wants the action to match $\theta$. Her payoff is $u_R(\theta,a) = - (a - \theta)^2$. Given bias $b \in [0, \infty)$, the sender wants the action of the receiver to match $\theta+b$. Thus, his payoff is ${u_S(\theta,a) = - (a - \theta - b)^2}$. The bias parameter measures their divergence of interest. 

Following CS, a (Bayes) Nash equilibrium is a family of type-conditional probability distributions over messages for the sender and a choice of action conditional on the message for the receiver, such that there is no profitable deviation by the receiver or by any type of sender. Babbling equilibria always exists, where the sender's message conveys no relevant information on the type and the receiver plays her ex-ante optimal action for messages that are sent with positive probability. 

We let two independent reinforcement learning agents play, as sender and receiver, the discretised cheap talk game. To allow learning, the two agents play the game multiple times, up to a maximum of $T=10^7$ periods. Both are programmed to take an action conditional on a state, first the sender and then the receiver. In each period, a state for the sender is a type drawn from $\Theta$ according to $p$, independently of previous interactions. Then, the sender takes an action from $M$, which represents the state for the receiver. Finally, the receiver takes an action from $A$ and agents collect their rewards.  Because the underlying learning model is the same for both agents (i.e., both take an action conditional on a state), we describe it for a generic agent, with states and actions taking appropriate meaning based on who is playing.

Let $\mathcal{S}$ be the finite set of possible states and $\mathcal{A}$ the finite set of actions, for either the sender or the receiver. Each time $t \in \{1, 2, \ldots, T\}$ an agent is called to play in state $s \in \mathcal{S}$, it chooses action  $a \in \mathcal{A}$ following a parameterised softmax probability distribution
\begin{equation*}\label{eq:softmax}
\pi_t(a \mid s) = \dfrac{ e^{ Q_t(s,a) / \tau_t}}{\sum_{a^\prime \in \mathcal{A}} e^{ Q_t(s,a^\prime) / \tau_t}},
\end{equation*}
where $Q_t(s,a)$ (discussed in the next paragraph) represents the agent's estimate in period $t$ of the value of taking action $a$ in state $s$. The parameter $\tau_t$, called temperature, modulates the intensity of exploration: for smaller values of $\tau_t$, the probability mass increasingly concentrates on the action(s) that are most rewarding according to $Q_t(s,a)$. We reduce exploration at each interaction by letting the temperature decay according to $\tau_t = \tau_{1} e^{-\lambda (t-1)}$, where $\lambda \in [0,1)$ is the decay rate and $\tau_1 = 0.1$. Hence, exploration goes to zero as $t \to \infty$.

The initial estimate, $Q_0(s,a)$, is  arbitrarily initialised for all $(s,a) \in \mathcal{S} \times \mathcal{A}$. If the agent takes action $a$ in state $s$ in period $t$, the estimate associated with that specific state-action pair is updated iteratively according to
\begin{equation*}\label{eq:update-rule}
Q_t(s,a) = Q_{t-1}(s,a) + \alpha \left[ r_t (s,a) - Q_{t-1}(s,a)\right],
\end{equation*}
where the step-size parameter $\alpha \in (0,1]$, called learning rate, regulates how quickly new information replaces the old and $r_t (s,a)$ (discussed later) denotes the reward the agent obtains by playing action $a$ in state $s$ in period $t$. For all other $(s',a')$ pairs, $Q_t(s',a')=Q_{t-1}(s',a')$.\footnote{The specification we adopt agrees with \citet{BanchioMantegazza22}'s definition of a reinforcer but differs from their $\epsilon$-greedy Q-learning in two ways. First, rather than a greedy policy, we employ softmax, which smooths out the effects of minor differences in the agents' estimates during learning. Second, because actions have no direct effect on future states, our players are not designed to estimate the long-run benefit of taking an action today. Formally, we set their discount parameter $\gamma$ to zero.} 

In multi-agent reinforcement learning, the rewards obtained in each period depend on the actions taken by other agents in that same period. In our case, let $(a',s')$ be the pair of state and action taken by the other agent in $t$. Then, we have $r_t (s,a)= - (a' - s - b)^2$ for the sender's algorithm and  $r_t (s,a)= - (a- s')^2$ for the receiver.

If the distribution of $r_t$ were to depend only on the agent's own actions, existing results would guarantee convergence of the policy $\pi_t(\, \cdot \mid s)$ to an optimal one. However, because the underlying distributions of rewards the agents face are non-stationary, convergence is not guaranteed.  For this reason, we consider agents to have converged and stop the simulation if, before reaching the maximum number of interactions $T$, the policies of both agents exhibit relative deviations in $L_{2,2}$ norm smaller than $0.1 \%$ for $10000  < T$ consecutive interactions. 

Pseudocode for the simulation is given in \autoref{al:pseudocode}.

\begin{algorithm}[!h]
	\begin{algorithmic}
		\medskip
		\State Initialise $Q^S$ and $Q^R$ arbitrarily
		\For{ each episode }
		\State $\theta \sim p(\theta)$
		\State $m \sim \pi^S(m \mid \theta)$
		\State $a \sim \pi^R(a \mid m)$	
		\State $Q^S(\theta, m) \gets  Q^S(\theta, m) + \alpha [u_S(\theta,a) - Q^S(\theta, m)]$
		\State$Q^R(m, a) \gets Q^R(m, a) + \alpha [u_R(\theta,a) - Q^R(m, a)]$
		\State $\tau \gets \tau e^{-\lambda}$
		\State  \textbf{if} $\pi^S$ and $\pi^R$ converged \textbf{then break}
		\EndFor
		\medskip
	\end{algorithmic}
	\caption{Independent reinforcement learning in the discretised cheap talk game}
	\label{al:pseudocode}
\end{algorithm}

\section{Main Results}\label{sec:results}

In this section, we discuss results from the baseline simulations, which involve a uniform prior and quadratic utility. The robustness of our findings to alternative game forms is demonstrated in the next section.

For our baseline case, we consider the discretised game with $n=6$ types in $[0, 1]$, with any two adjacent types separated by a $0.2$ increment. Hence, $\Theta = \{0.0, 0.2, 0.4, 0.6, 0.8, 1.0\}$, $M=\Theta$ and $A = \{0.0, 0.1, \ldots, 0.9, 1.0\}$.

We implement algorithms for both the sender and the receiver that use the same learning rate $\alpha = 0.1$ and exploration decay-rate $\lambda = 5 \times 10^{-6}$. With these hyperparameters, the agents typically converge in less than one million periods, and rewards' weight in the estimates is less than $1\%$ after $23$ updates. We test robustness to different hyperparameter configurations at the end of this section. The Q-matrices of the sender and of the receiver have dimensions $6 \times 6$ and $6 \times 11$, respectively. Their entries are initialised using a uniform distribution in the interval $[-7/60 -b^2,0]$ for the sender and $[-7/60,0]$ for the receiver, where the lower bounds correspond to the payoffs the two agents obtain ex-ante in the babbling equilibrium. 
Additional simulations confirmed that the initialisation of the matrices is irrelevant.

We study interactions for different levels of bias, taking $101$ points spaced $0.005$ apart in the interval $[0,0.5]$. For each bias $b$ we run 1000 independent simulations. At the end of each simulation, if the agents' policies have converged, we record the Q-matrices at the point of convergence and compute the implied policies for the sender and receiver, denoted $\pi_{\infty}^S(\, \cdot \mid \theta)$ and $\pi_{\infty}^R(\, \cdot \mid m)$, respectively. Using these policies we can compute the ex-ante expected rewards of the agents from playing the information transmission game together:
\begin{align*}
U_S ( \pi_{\infty}^S,  \pi_{\infty}^R) &= -\sum_{\theta}  p(\theta) \sum_m \pi_{\infty}^S(m \mid \theta) \sum_a  \pi_{\infty}^R(a \mid m)(a-\theta-b)^2,\\[5pt]
U_R  ( \pi_{\infty}^S,  \pi_{\infty}^R)  &= -\sum_{\theta}   p(\theta) \sum_m  \pi_{\infty}^S(m \mid \theta)\sum_a  \pi_{\infty}^R(a \mid m)(a-\theta)^2.
\end{align*}
We also compute the mutual information between type and message, normalised by the entropy of the type. This is equivalent to the expected relative reduction in the entropy of the type from knowing the message. Formally,
 \begin{equation*}\label{eq:mutual-information}
 I( \pi_{\infty}^S) = \left(\sum_{\theta} p(\theta) \log\left(\frac{1}{p(\theta)}\right)\right)^{-1} \sum_{\theta}\sum_{m} \pi_\infty^S(m \mid \theta) p( \theta) \log\left(\frac{ \pi_\infty^S(m \mid \theta)}{\sum_{\theta} \pi^S_\infty(m \mid \theta) p(\theta)}\right).
 \end{equation*}
 This metric takes value 1 if knowledge of the message implies knowledge of the type, as in a perfectly informative equilibrium. It takes value 0 when type and message are statistically independent, as in the babbling equilibrium. Other measures of informativeness do not have qualitatively different behaviour. In particular, in the uniform-quadratic case, the negative of the residual variance of the sender's type given the message coincides with the ex-ante payoff of the receiver, $U_R$.
 
\subsection{Convergence to Nash}
\begin{figure}[H]
	\center
	\includegraphics{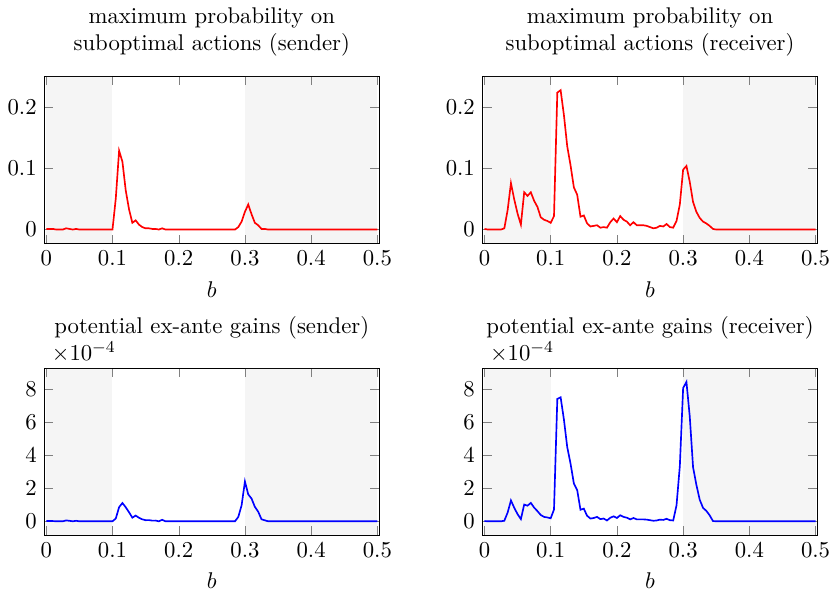}
	\caption{Top: Maximum probability mass that the policy of the sender (receiver) places on suboptimal messages (actions) across all types (and messages). Bottom: Potential ex-ante gain by a unilateral deviation across all types (and messages). Averages over $1000$ simulations.\\
	\textit{Also applies to subsequent figures}: The ex-ante optimal equilibrium entails perfect information transmission for biases identified by the shaded grey area to the left, while babbling is the unique equilibrium for biases in the shaded grey areas to the right.}
	\label{fig:base-case-error}
\end{figure}

In our baseline case, all simulations converged according to the stated criterion and, as the next figure illustrates, most converged to a (Bayes) Nash equilibrium. The top panel of Figure \ref{fig:base-case-error}, shows the maximum probability mass, across all states, that the policy of the sender (and that of the receiver) places on actions that are not best responses to the policy of the opponent. The bottom panel of Figure \ref{fig:base-case-error} displays the ex-ante gain the sender and the receiver would make by best responding to the policy of the opponent, thus measuring epsilon equilibrium behaviour \citep{Radner1980}. 

Except at a few bias levels around where the perfectly informative equilibrium disappears and where babbling becomes the unique equilibrium, the agents converge to an exact Nash with high precision. When they do not, their loss from not best responding is low, on average around 0.00085 at most. While an explicit discussion is omitted from the next section, we note that this result remains valid for different specifications of the game form.

To evaluate how the choice of hyperparameters affects this conclusion, we run our simulations of the cheap talk game for a grid of reinforcement learning hyperparameters. We consider $\alpha \in \{0.025,0.05,0.1,0.2,0.4\}$ and $\lambda \in \{2, 1, 0.5 , 0.25, 0.125\} \times 10^{-5}$. The exploration decay rates are spaced so that the number of periods to converge roughly doubles at each step.\footnote{In practice, with $\lambda = 2 \times 10^{-5}$ it takes approximately $2 \times  10^5$ interactions for the agents' policies to converge, and with $\lambda = 1.25 \times 10^{-6}$ it takes approximately $3.2 \times 10^6$ interactions.} In Figure \ref{fig:grid-search-heat-90quant-absolute-error} below, we report the frequency of simulations where agents' policies converged close to a Nash equilibrium.

\begin{figure}[H]
	\center
	\includegraphics{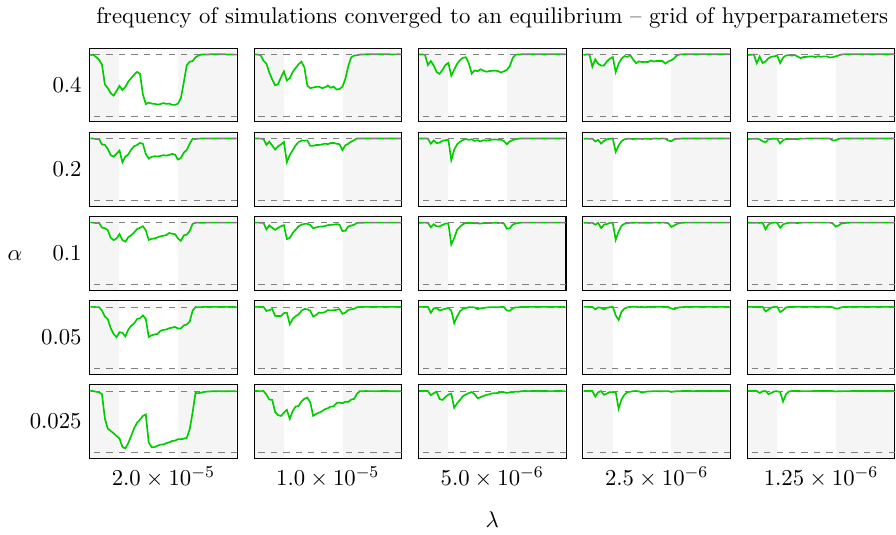}
	\caption{Frequency of simulations in which both agents place at most $0.01$ probability mass on suboptimal actions across all states, for different levels of bias in \([0,0.5]\). Each graph has bias on the horizontal axis and frequency on the vertical axis, and corresponds to a specific $(\lambda, \alpha)$ combination of hyperparameters. Horizontal dashed lines indicate frequencies of 0 and 1. }
	\label{fig:grid-search-heat-90quant-absolute-error}
\end{figure}

Figure \ref{fig:grid-search-heat-90quant-absolute-error} indicates that, while intermediate learning rates are most favourable for convergence to equilibrium, letting the agents explore more by reducing the learning decay rate $\lambda$ has an unambiguously positive effect on convergence to equilibrium. 

\subsection{Equilibrium Selection}

The CS game of information transmission has multiple Nash equilibria and this remains the case in its discretised version. While babbling is the unique equilibrium when the bias is above 0.3, other equilibria that induce different levels of information transmission exist with lower bias. The existing literature does not offer a complete characterisation of such equilibria in the discretised model. \citet{Frug16} (Proposition 2) shows that in the uniform-quadratic case the set of Pareto efficient equilibria is a singleton, payoff-wise. This equilibrium, which \citet{Frug16} characterises, represents an important benchmark for us and is henceforth referred to as ``optimal''.

One class of equilibria, which we refer to as monotone partitional equilibria, is especially relevant to study equilibrium selection because our simulations nearly always converge to strategies in such class. In a monotone partitional equilibrium, the sender partitions $\Theta$ into contiguous intervals and uses an (ex-ante) strategy which is measurable with respect to the partition. As we shall see, this class is never a singleton in our specifications, except in cases where babbling is the unique equilibrium. In fact, both the babbling equilibrium and the optimal equilibrium characterised in \citet{Frug16} are monotone partitional.\footnote{\citet{Frug16} endows the receiver with a continuum of actions. This ensures that a single optimal action corresponds to each belief the receiver may have. Given our discretisation of $\Theta$ and $A$, the optimal action is unique as long as the strategy of the sender is monotone partitional. The optimal equilibrium constructed in \citet{Frug16} is also optimal in our setting because in it the strategy of the sender is partitional.} To fix ideas, the modal policies at convergence for both the sender and the receiver are illustrated below for some selected levels of the bias.

\begin{figure}[H]
	\center
	\includegraphics{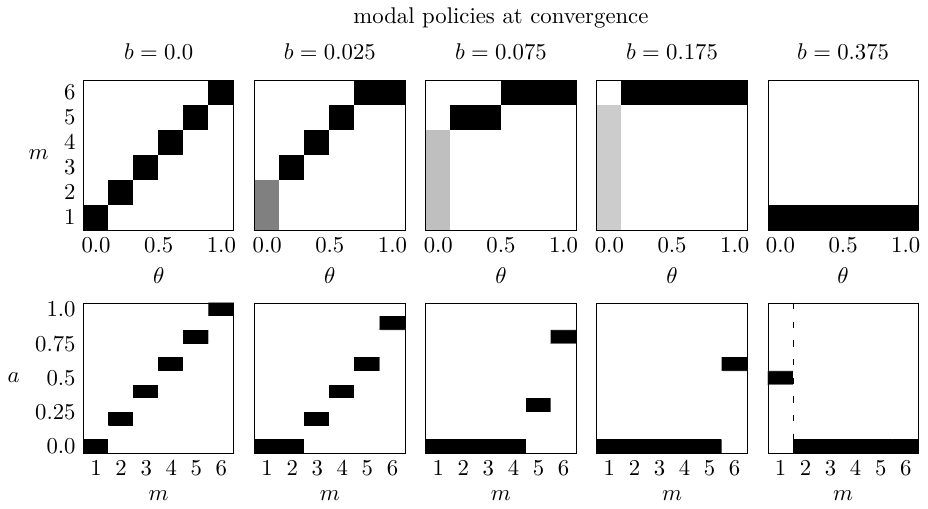}
	\caption{Heathmap of the modal policies of sender (top) and receiver (top) for different levels of bias over 1000 independent simulations. All vertical pairs of strategies correspond to exact equilibria. Randomisation over messages is with equal probability as indicated by the same colour tone. Messages to the right of the dashed line are off the equilibrium path.  To find the modal policy we relabeled messages in each simulation assigning a natural number to each message such that messages with smaller numbers are associated with smaller types.}
	\label{fig:base-case-group-policies}
\end{figure}

Having established that reinforcement learning dynamics prevalently lead to partitional Nash and that the game has many such equilibria exhibiting different levels of informativeness, we now study where our algorithms converge as a function of the bias.  

\begin{figure}[H]
	\center
	\includegraphics{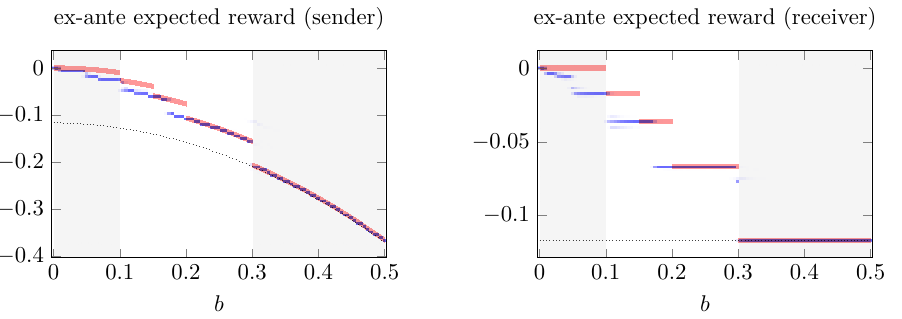}
	\includegraphics{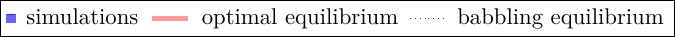}
	\caption{Ex-ante expected reward for the sender (left) and receiver (right) for different levels of bias. The distribution of values of 1000 simulations is shown in shades of blue. The value associated with the ex-ante optimal equilibrium is in red and the one associated with the babbling equilibrium is dotted gray.}
	\label{fig:base-case-rewards}
\end{figure}

\begin{figure}[H]
	\centering
	\hspace{4.6pt}	
	\begin{minipage}[t]{0.48\textwidth}
		\center
		\includegraphics{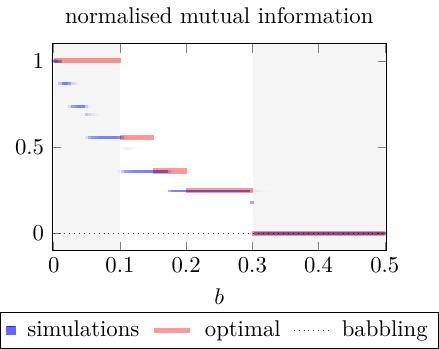}
		\captionsetup{margin=0cm}
		\caption{Normalised mutual information between the distribution of messages induced by the sender's policy and the distribution of sender's types. The distribution over 1000 simulations is shown in shades of blue. The value associated with the optimal equilibrium is in red and the one associated with the worst equilibrium is dotted gray.  }
		\label{fig:base-case-mutual-information}
	\end{minipage}
	\hfill
	\begin{minipage}[t]{0.48\textwidth}
		\center
		\vspace{-167.8pt}	
		\includegraphics{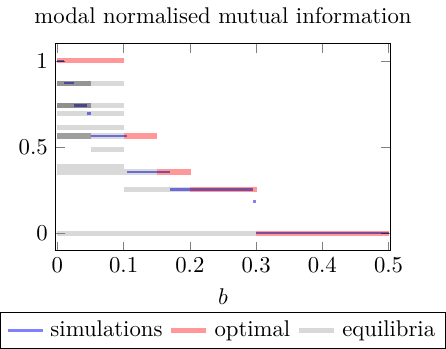}
		\captionsetup{margin=0cm}
		\caption{Normalised mutual information of the sender's modal policy across simulations converged to an equilibrium (maximum mass on suboptimal actions across states < 0.01 for both agents). The normalised mutual information of monotone partitional equilibria that exist for a given bias is shown in grey.}
		\label{fig:baseline_modal_mutual_information}
	\end{minipage}
\end{figure}

In \autoref{fig:base-case-rewards}, we compare the distribution (blue heatmap) of ex-ante payoffs arising from the simulations to the theoretical bounds provided by the babbling equilibrium (dotted line) and the optimal equilibrium (in red) for the different levels of bias in the discretised  $[0,0.5]$ interval. In \autoref{fig:base-case-mutual-information}, we show the distribution of the normalised mutual information between type and message from our simulations, for the same range of biases. Finally, in \autoref{fig:baseline_modal_mutual_information} we identify all monotone partitional equilibria that exist in our discretised game at different levels of the bias, distinguishing them by their level of informativeness. Equilibria where the algorithms converge are highlighted in blue.

Two conclusions are immediate and, as we shall see in the next section, robust to different specifications of the game. First, unless babbling is the only equilibrium, our agents converge to play with substantial information transmission,  which in this model is reflected both by the mutual information and agents' payoffs above the babbling level. Second, the level of information transmission that takes place at convergence decreases monotonically as the bias increases, which seems a natural property of any equilibrium selection criterion in the information transmission game. 

Albeit requiring a more granular observation of the data, a clear pattern also emerges at levels of the bias where neither perfect information transmission is an equilibrium nor babbling is the unique equilibrium, that is between 0.1 and 0.3 bias in this baseline case. In particular, play converges to either the optimal equilibrium or to the second-best one, which becomes the optimal one at higher levels of the bias. Behaviour is more nuanced at lower levels of bias. In this case, as \autoref{fig:baseline_modal_mutual_information} shows, the game exhibits many equilibria which then disappear once 0.1 is reached. The agents only play the perfectly informative equilibrium when the bias is nearly zero. Then, as the bias grows toward 0.1, they gradually descend into less and less informative equilibria, reaching the Nash that is optimal at $>0.1$ bias midway at around $0.05$. As the left panel of \autoref{fig:base-case-rewards} shows, the transition is smoother for the sender, suggesting that jumps from one equilibrium to the other take place when the seller becomes nearly indifferent between the two equilibria. 

The behaviour described in this subsection is robust to alternative specifications of the learning hyperparameters. \autoref{fig:grid-mutual-information} below shows the distribution of the normalised mutual information for values of $\alpha \in \{0.025,0.1,0.4\}$ and $\lambda \in \{2, 0.5, 0.125\} \times 10^{-5}$.

\begin{figure}[H]
	\center
	\hspace{-20pt}
	\includegraphics{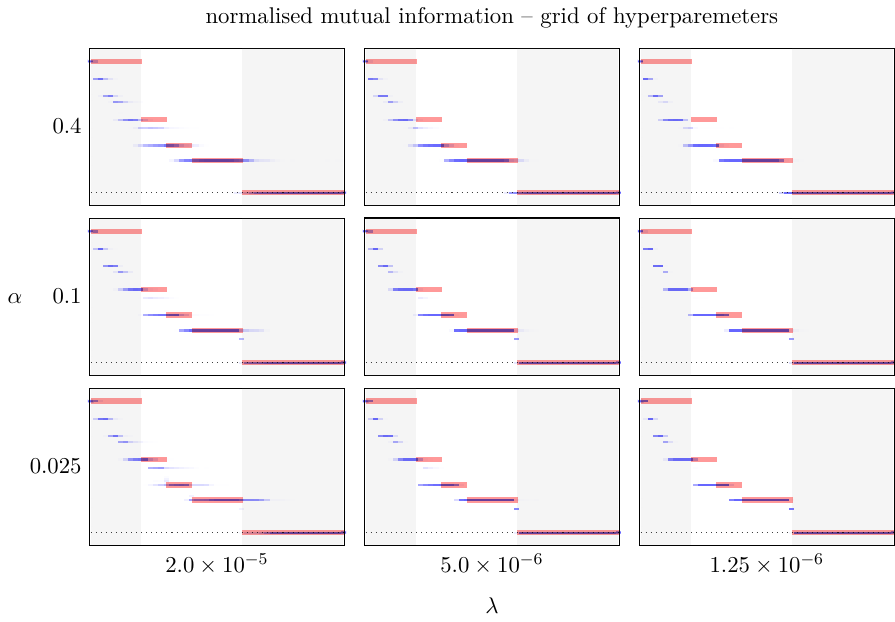}
	\includegraphics{figures/tikz_externalize/legend_baseline_expected_rewards.pdf}
	\caption{Normalised mutual information for a grid of hyperparameters; as in \autoref{fig:base-case-mutual-information}. Distribution of outcomes over $1000$ simulations.}
	\label{fig:grid-mutual-information}
\end{figure}

In addition to showing robustness of the results, \autoref{fig:grid-mutual-information} highlights that letting agents explore more extensively yields outcomes that are progressively closer to the ex-ante optimal equilibrium. The same trend naturally extends to the normalised mutual information between messages and types.

\section{Comparative statics and Robustness}\label{sec:robustness}

In this section, we demonstrate that communication emerges robustly in information transmission games played by AI agents, beyond the classic uniform-quadratic case studied so far. First, we look at making the language more or less expressive than it is necessary to achieve all equilibria of the game and, analogously, the set of actions more or less large. Second, we report the results of simulations obtained for a variety of alternative assumptions on the information transmission game. We consider a higher and lower number of types, non-uniform prior, and utility functions that are not linear-quadratic.  Throughout, we keep fixed the reinforcement learning hyperparameters as in our baseline configuration. We show for each case how the ex-ante expected reward of the agents is distributed over 1000 simulations. 

\subsection{Comparative statics on messages and actions}
In this subsection, we perform some comparative statics by changing the number of messages available to the sender and the number of actions available to the receiver.

\begin{figure}[H]
	\center
	\includegraphics{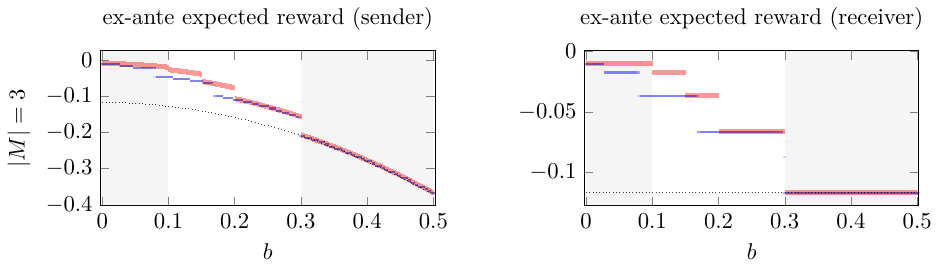}\\[2pt]
	\includegraphics{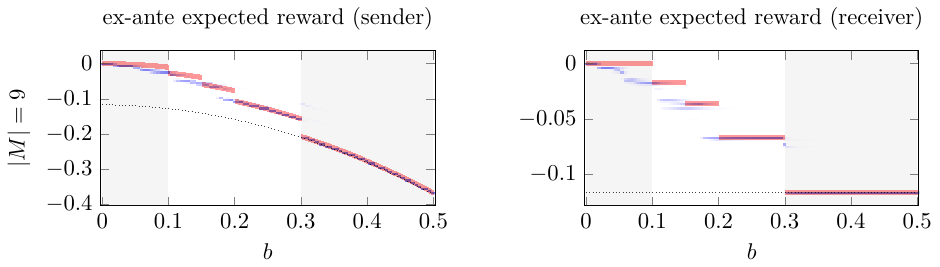}
	\includegraphics{figures/tikz_externalize/legend_baseline_expected_rewards.pdf}
	\caption{Ex-ante expected reward for the sender (left) and receiver (right) for different levels of bias. Cases with $3$ messages (top) and $9$ messages (bottom).}
	\label{fig:robustness-nmessages-rewards}
	\vspace{2pt}
	\center
	\includegraphics{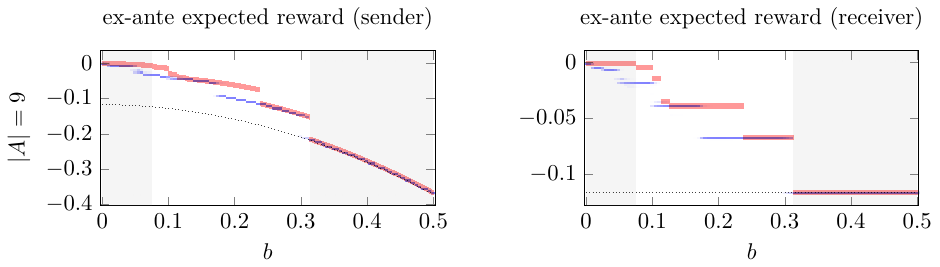}\\[2pt]
	\includegraphics{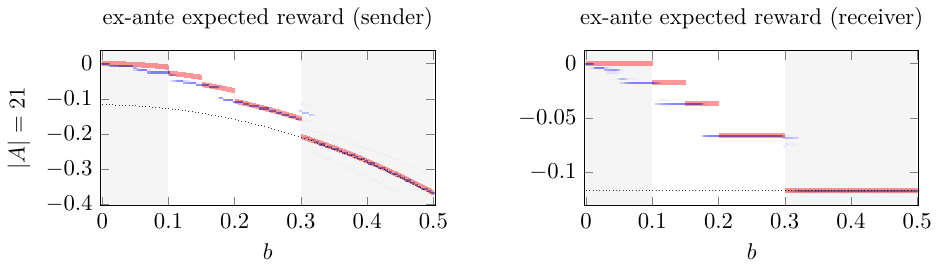}
	\includegraphics{figures/tikz_externalize/legend_baseline_expected_rewards.pdf}
	\caption{Ex-ante expected reward for the sender (left) and receiver (right) for different levels of bias. Cases with $9$ actions (top) and $21$ actions (bottom).}
	\label{fig:robustness-nactions-rewards}
\end{figure}

In  \autoref{fig:robustness-nmessages-rewards} we endow the sender with more or fewer messages than the number required to achieve full information transmission. In particular, while keeping fixed the number of types to 6, we look at the case where only 3 messages are available and the case where 9 messages are present. Results are in line with expectations. On the one hand, increasing the number of messages does not have a tangible effect on the qualitative conclusions reached in the previous section. The sender simply learns to avoid redundant messages and the receiver responds to those in a way that does not stimulate the sender to play those actions. Learning becomes slower though, because the dimensions of the Q-matrices increase. On the other hand, making the language less expressive may impede players from playing the more informative equilibria. In fact, when the constraint on messages is non-binding, behaviour is as in the baseline scenario and less noisy because less learning is required. Instead, when the constraint is binding, sender and receiver tend to set on the most informative equilibrium compatible with the number of messages available. 

In  \autoref{fig:robustness-nactions-rewards} we modify our baseline scenario by increasing or reducing the number of actions available to the receiver compared to the baseline case of 11. In particular, we consider cases with $9$ and $21$ uniformly spaced actions in the unit interval. Both are odd numbers, which guarantee the set of actions to contain the best reply to babbling, which is 0.5. As in the case of superfluous messages, if the number of actions is large enough to include all those required in the equilibrium or more, there is no substantial change in behaviour at convergence compared to the baseline scenario. The main qualitative conclusions we reached in the baseline scenario also continue to hold when the number of actions available to the receiver is reduced. However, for any number of actions smaller than $11$, the best reply to partitional strategies may not be unique and the ex-ante optimal equilibrium, hence our benchmark, is different than the one identified by \citep{Frug16}.

\subsection{Robustness to alternative game forms}
We now look at specifications with different numbers of types, different utility specifications and different distributions over the types.

\begin{figure}[H]
	\center
	\includegraphics{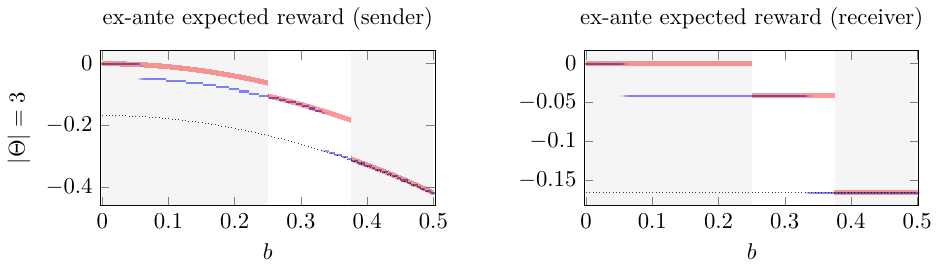}\\[2pt]
	\includegraphics{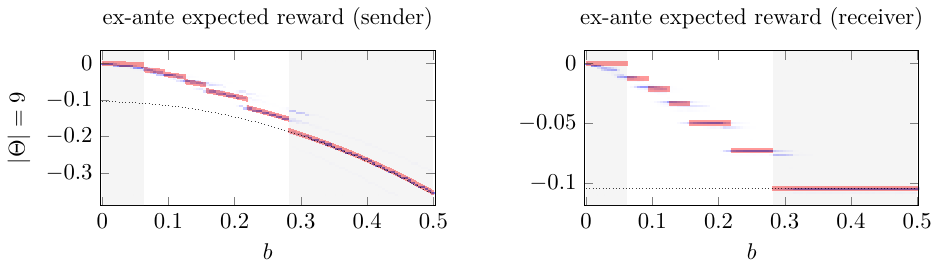}
	\includegraphics{figures/tikz_externalize/legend_baseline_expected_rewards.pdf}
	\caption{Ex-ante expected reward for the sender (left) and receiver (right) for different levels of bias. Cases with $3$ types (top) and $9$ types (bottom).}
	\label{fig:robustness-nstates-rewards}
	\vspace{2pt}
	\center
	\includegraphics{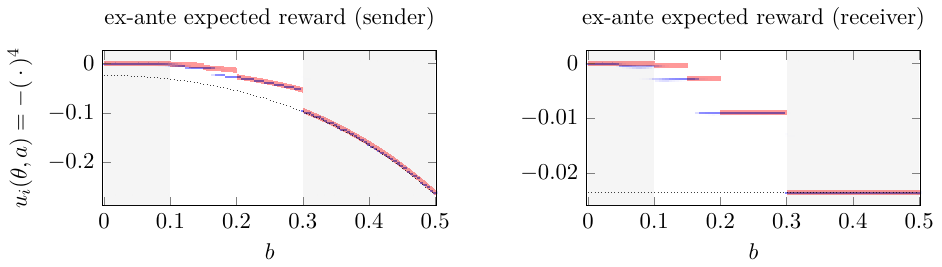}\\[2pt]
	\includegraphics{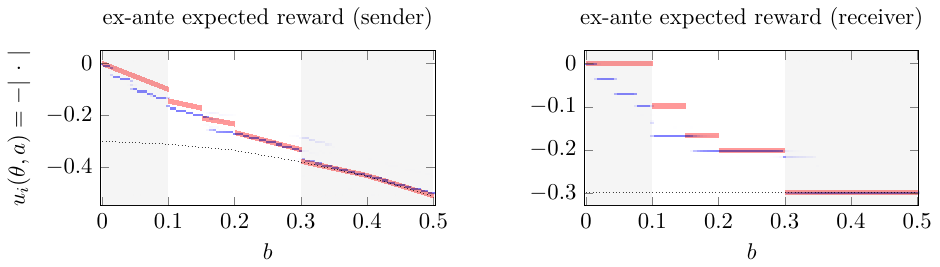}
	\includegraphics{figures/tikz_externalize/legend_baseline_expected_rewards.pdf}
	\caption{Ex-ante expected reward for the sender (left) and receiver (right). Fourth-power loss (top) and absolute loss (middle).}
	\label{fig:robustness-utilities-rewards}
\end{figure}

In \autoref{fig:robustness-nstates-rewards} we consider simulations with $n=3$ and $n=9$ types, so that any two adjacent types are spaced $0.5$ and $0.125$ from each other, respectively. The figure confirms that all conclusions reached in the previous section extend to the case of a smaller or larger type space. However, with a large number of types, behaviour tends to be more noisy. This is explained by the relative decrease in exploration due to the change in size of the agents' Q-matrices.  In general, as we keep $\lambda$ fixed to the base case configuration, each state-action pair is on average visited more (less) often the smaller (larger) the agent's Q-matrix is. This eventually results in improving (worsening) the agent's learning. 

\autoref{fig:robustness-utilities-rewards} shows simulation outcomes with different utility specifications. We consider the case of a fourth-power loss function and the case of an absolute loss function. Both functions are concave and, together with the rest of our assumptions, this guarantees the existence of an optimal monotone partitional equilibrium. Then, the figure confirms that our main results are not dependent on the specific forms of the utility function. Both scenarios show similar results, in line with our benchmark case. While we have not run further cases because it is hard to identify the right comparator, we strongly suppose that the assumptions of concavity and upward bias are not crucial for the main result that communication will take place at the highest levels predicted by equilibrium.

\begin{figure}[H]
	\center
	\includegraphics{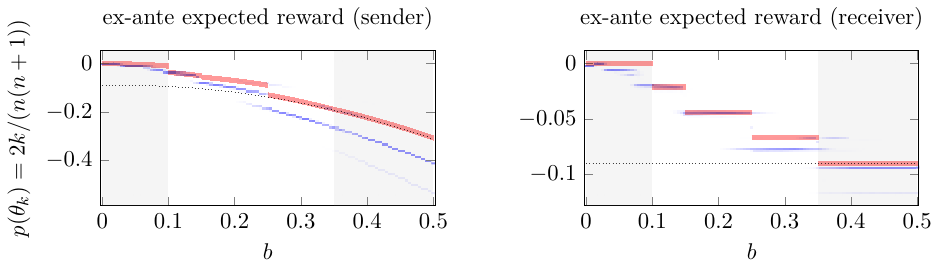}\\[2pt]
	\includegraphics{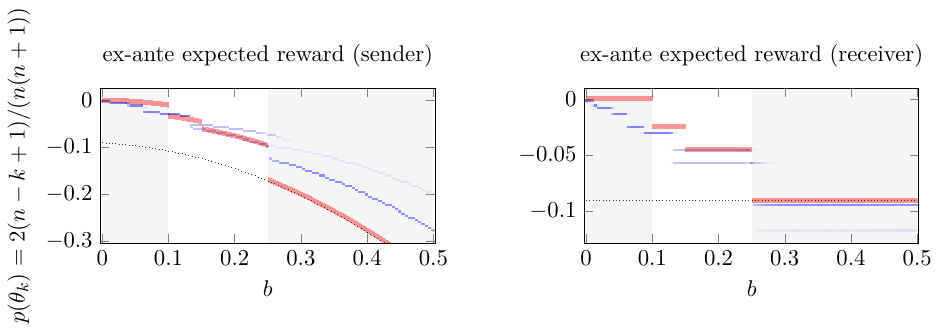}
	\includegraphics{figures/tikz_externalize/legend_baseline_expected_rewards.pdf}
	\caption{Ex-ante expected reward for the sender (left) and receiver (right) for different levels of bias. Cases with a linearly increasing distribution (top) and linearly decreasing distribution (bottom). We use $p(\theta_k)$ to indicate the probability mass on the $k$-th type in $\Theta=\{\theta_1,\ldots,\theta_n\}$. There are $n=6$ types as in the base-case simulations.}
	\label{fig:robustness-dists-rewards}
\end{figure}

Finally, in \autoref{fig:robustness-dists-rewards} we show outcomes for different distributions over the types; namely, a probability distribution with linearly increasing probability mass, and one with linearly decreasing probability mass. In this case results also indicate communication is roughly in line with the most optimistic theoretical benchmark. Now, however, existence of the ex-ante optimal equilibrium is no longer guaranteed. Hence, we rely on the receiver-preferred partitional equilibrium as comparator.  A surprising result is obtained in the case of the decreasing distribution. While in all our simulations agents do better than babbling, here the receiver obtains a payoff lower than the babbling one, while the sender obtains a larger one. We think this finding is interesting because the sender seems able to manipulate the receiver even when theoretically it should not be possible and the receiver is losing out from not just playing the ex-ante optimal action. Unfortunately, we do not have a convincing explanation for this result. We believe it might be due to the existence of multiple optimal actions for the receiver even when the sender’s strategy is partitional.

\section{Conclusions}\label{sec:conclusion}
We showed that simple reinforcement learning algorithms training together in the classic \citet{CrawfordSobel82} cheap talk game engage in substantial information transmission at the level predicted by the most informative equilibrium.  

In the uniform-quadratic version of CS, equilibria exhibit a nice structure. Both sender and receiver unambiguously benefit from more communication. This raises the question of what would happen in games with multiple equilibria that are not Pareto ranked, with some more favourable to the receiver and others to the sender. Will communication break down? Or will one of the two agents lead the other to their favourite equilibrium? While our results in Section 4 suggest that communication will persist and favour the sender, we believe extending the analysis to more general games with communication is an interesting avenue for future work.

Our results are interpreted as arising from a large population of randomly matched agents. A natural extension of the present framework, which would require explicitly simulating the population dynamics, would be looking at how a language is learned when agents within the populations are heterogeneous (e.g., senders may have different biases) or the frequency of interactions is not driven by random matching (e.g., agents may be arranged in a network as in \citet{skyrms2009}'s model of unbiased communication). Would agents still be able to learn a common language? Will there be winners and losers depending on the level of bias or the network architecture of interactions? What population structures facilitate learning?

A more speculative next step would be to consider the interaction between humans and algorithms. Suppose we let a multitude of humans randomly interact with multiple algorithms. Will they learn a common language? Will there be more communication than in human-to-human experiments? Would humans be manipulated or maybe the other way around? Human-algorithm play also raises interesting questions regarding the interaction between strategic signalling and natural language. How would reinforcement learning algorithms endowed with natural language processing abilities, such as those currently possessed by chatGPT and other large language models, perform? Will the use of natural language result in more or less information transmission? Will human agents be more easily deceived? We think that human-AI experiments show promise well beyond the questions raised above.

Finally, it may be worth revisiting some of the existing findings in the economics of AI agents playing market games. For instance, will the sort of code-bidding collusion described in the introduction emerge in market played by AI agents? Since a large state space would be required to handle this sort of ``non-verbal'' exchange, our finding that communication emerges suggests it may be worth looking at the behaviour of more complex agents, such as those endowed with deep neural networks. It would not be surprising to see collusion sustained at higher levels than those already observed with simple learning algorithms. Such a finding would suggest the need for market design to mitigate communication possibilities. 

\newpage
\bibliographystyle{apalike}
\bibliography{bibfile}

\end{document}